\documentstyle[twocolumn,aps]{revtex}
\newcommand{\beqn}{\begin{eqnarray}}
\newcommand{\eeqn}{\end{eqnarray}}
\renewcommand{\theequation}{\thesection.\arabic{equation}}
\newcommand{\sss}{\scriptscriptstyle}
\begin{document}

\title{ Some Remarks on the Quantization of Gauge Theories}

\author{ Radhika Vathsan \thanks{ email: radhika@imsc.ernet.in.} }

\address{The Institute of Mathematical Sciences,
C.I.T. campus, Taramani, Madras 600 113, India.}


\maketitle

\begin{abstract}
  The methods of reduced phase space quantization and Dirac quantization
are examined in a simple gauge theory. A condition for the possible
equivalence of the two methods is discussed.



\end{abstract}

\section*{Introduction}

    A gauge theory is regarded in the canonical framework as a system
with first class constraints \cite{Dirac}. In the classical analysis
according to Dirac, the Hamiltonian $H$ is the canonical one $H_c$
plus an arbitrary linear combination of the first class constraints
 $\phi_i$.
This means that the classical trajectories involve arbitrary
functions of time: the Lagrange multipliers $\lambda_i$. So
a given physical state doesn't correspond to a unique set of
canonical variables on the phase space $\Gamma$. This problem can be
circumvented in either of two ways:
\begin{itemize}
\item Gauge fixing constraints $\chi_i$ are introduced, one
	for each $\phi_i$, such that they are preserved in
	time, i.e.
		\[ \{H,\chi_i\}\simeq 0 \]
	and the matrix $ C_{ij} = \{\phi_i,\chi_j\}$ is nonsingular.
	(This then becomes a theory with second class constraints.)
	Thus the $\lambda_i$'s are fixed so that evolutions from initial
	states on the submanifold $\Gamma^*$ defined by $\phi_i=\chi_i = 0$
	are unique. (For future reference, we will denote by ${\Gamma^*}'$
	the constraint surface $\phi_i=\chi_i = 0$ and provided det$C \neq 0$
	everywhere on the surface we will denote it by $\Gamma^*$ and
	refer to it as the reduced phase space.)

\item Since the $\lambda_i$'s bring in the arbitrary time dependence,
	 all points on an orbit $\cal O$ generated by the gauge
	generators $\phi_i$ must be regarded as physically equivalent.
	So if $\hat{\Gamma}$ is the constraint surface $\phi_i = 0$,
	then the true dynamical trajectories lie on
	$\tilde{\Gamma}\equiv \hat{\Gamma} / \sim $, where $\sim$ is
	the equivalence relation $P\sim P'$ if $P, P' \in \cal O$.
\end{itemize}
The surface $\Gamma^*$ is diffeomorphic to $\tilde{\Gamma}$ provided
the surface $\chi_i = 0 $ intersects each orbit in $\hat{\Gamma}$ exactly
once. This condition on the gauge-fixing constraints
$\chi_i$ is a prerequisite for the equivalence of the two approaches.
For the first case, the condition of invertibility of the matrix
$C_{ij}$ ensures that {\em{locally}} the $\chi_i = 0 $ surface
intersects $\hat{\Gamma}$ only once, but not necessarily globally.
This point has bearing on the quantization of a gauge theory,
since quantum theory is sensitive to the global properties of
the phase space to be quantized.

 These two approaches have their counterparts in the quantization of gauge
theories --
\vspace{2mm}
\newline
{\bf Method A:}  Reduced Phase Space Quantization--
 fix the gauge to obtain the space $\Gamma^*$ and define the
Poisson bracket structure on this as the Dirac
 brackets on the original phase space $\Gamma$. $\Gamma^*$ so equipped
is called the reduced phase space. It can then be directly quantized,
which involves the finding of a commutator
algebra representation for Poisson brackets. (This process can
be complicated because the reduced phase space $\Gamma^*$
is not always topologically trivial.) So here one quantizes {\em after}
reducing the phase space.
\vspace{2mm}
 \newline
{\bf Method B:}  Dirac Quantization-- canonically quantize the original
phase space $\Gamma$ (which is usually ${\bf R}^{2N}$)
and then impose the gauge constraints as operator conditions on
the physical quantum states: \[\hat{\phi_i} \psi_{phys} =0.\] These are
sometimes referred to as supplementary conditions. This
is quantization {\em before} reduction.

Notice that method A depends manifestly on a choice of gauge-fixing
constraints $\chi_i$ and there is a vast freedom in this choice, in general.
An immediate question is whether method A applied with two different
choices of the $\chi_i$'s gives equivalent quantum theories. Method B, on
the other hand, is manifestly independent of any choice of gauge. If the two
methods give equivalent quantum theories, then the manifest gauge-invariance
of method B reflects the gauge-independence of method A applied on a class
of $\chi_i$'s. The discussion of the possible equivalence of $\Gamma^*$ and
$\tilde{\Gamma}$ has a crucial role to play in the equivalence
of the quantum theories obtained by these two methods.

These matters are illustrated in the present work in the context of a very
simple toy model gauge theory.

The model considered is described in the first section of the paper. The
second section deals with its quantization by method A and the third section,
method B. The choice of constraints and a discussion of a condition for
the equivalence of these two methods is discussed in the fourth section.
A discussion of and conclusions from the lessons learnt from the
exercise comprise the fifth section. An appendix is included, giving a short
review of the geometric quantization technique used in the quantization of
the reduced phase space, along with the details of the calculations for the
present case.

\section{The Toy Model}

We consider the phase space ${\bf R}^4$ with canonical coordinates $q^{\sss 1},
q^{\sss 2}, p_{\sss 1}, p_{\sss 2}$ and the constraints:
\beqn
\phi \equiv {q^{\sss 1}}^2 + {q^{\sss 2}}^2 + p_{\sss 1}^2 + p_{\sss 2}^2 - R^2
=0 \label{c-1}
\eeqn and
\beqn
\chi \equiv p_{\sss 2} = 0 \label{c-2}
\eeqn
Suppose we regard the constraint $\phi$ as the gauge generator or
the first class constraint and $\chi$ as the gauge-fixing condition.
The constraint surface ${\Gamma^*}'$ is thus the 2-sphere $S^2$.
The matrix
\beqn
C\equiv \{\phi,\chi\} =\left( \begin{array}{cc}
                              0 & 2q^{\sss 2} \\
                             -2q^{\sss 2}& 0
                              \end{array}  \right)
\eeqn
                             is non-singular provided
$q^{\sss 2} \neq 0$. This immediately shows that reduced phase space cannot be
${\Gamma^*}'$. Let us proceed, nevertheless, and see how to obtain the true
reduced phase space $\Gamma^*$.

The Poisson bracket \{.,.\} on
${\bf R}^4$ must be modified to the Dirac bracket \{.,.\}* on
the constraint surface. This is given by
\beqn
 \{f,g\}^* = \{f,g\} - \sum_{i,j}\{f,\xi_i\}C_{ij}^{-1}\{\xi_j,g\}.
\eeqn
where  $\xi_i$ is a second class constraint  and
$f,g \in C^\infty({\bf R}^4)$.
The Dirac brackets of the canonical coordinates are
\beqn
    	\{q^{\sss 1},q^{\sss 2}\} &=& -{p_{\sss 1}\over q^{\sss 2}}\nonumber\\
	\{q^{\sss 1},p_{\sss 1}\} &=& 1 	\\ ~\label{db}
	\{q^{\sss 2},p_{\sss 1}\} &=& -{q^{\sss 1}\over q^{\sss 2}},\nonumber
   \eeqn
the rest being zero. Introducing the standard coordinates ($\theta',\varphi'$)
on the sphere $S^2 \sim {\Gamma^*}'$, $(q^i,p_i)$ can be parametrized as
 \beqn
 q^{\sss 1} &=& R\sin{\theta'}\cos{\varphi'} \nonumber\\
 p_{\sss 1} &=& R\sin{\theta'}\sin{\varphi'} \\
 q^{\sss 2} &=& R\cos{\theta'} \nonumber
\eeqn
where $0 \leq \theta' \leq \pi$ and $0 \leq \varphi' \leq 2\pi$.
This is singular at $\theta' = \pi/2$ which corresponds
to the singularity of the Dirac brackets (\ref{db}) at $q^{\sss 2} = 0$,
at the equator of the sphere.
The Dirac bracket, which is also the induced 2-form from ${\bf R}^4$, is
\beqn
 \{f,g\}^* = {1\over R^2 \sin{\theta'}\cos{\theta'}} \left({\partial{f}\over
\partial{\theta'}} {\partial{g}\over\partial{\varphi'}} -
{\partial{f}\over\partial{\varphi'}} {\partial{g}\over\partial{\theta'}}
\right) \label{DB}
\eeqn
defines a symplectic form on ${\Gamma^*}'$ minus the equator: the constraint
surface
${\Gamma^*}'$ is {\em not} the reduced phase space.
The reason for this, as shall be demonstrated below, is that the set of points
on ${\Gamma^*}'$ are not in 1--1 correspondence with the set of inequivalent
orbits of $\phi$ on the surface $\hat{\Gamma} \equiv \phi = 0$. Also, reduced
phase
space on which the above Dirac bracket defines a symplectic structure must be
obtained by a gauge-fixing condition that selects {\em one} point from
each orbit $\cal O$. The $\chi$ of eqn.(\ref{c-2}) does not satisfy this
criterion.
This is now shown explicitly.

The orbits $\cal O$ are the integral curves of the Hamiltonian vector fields
corresponding to $\phi$, which are described by the differential equations
\[ \dot{x} = \{x,\phi\}\]
\beqn
\Rightarrow \quad \dot{q}_i = 2p_i ,\quad \dot{p}_i = -2q_i, \quad i=1,2.
\eeqn
The general solution is
\beqn
q_i(t) = A_i \cos(2t - \alpha_i), \quad p_i(t) = -A_i \sin(2t - \alpha_i),
\eeqn
with $A_i > 0$, i.e. circles of radii $A_i$ in the $q_i$--$p_i$ planes, with
initial
conditions specified by the four parameters $(A_i,\alpha_i)$. Not all such sets
specify distinct orbits: if the set $(A_i,\alpha_i)$ lies on the orbit
generated from the set $(A_i',\alpha_i')$ then the two sets describe the
same orbit. This happens when
$A_i=A_i'\neq 0$ and $\alpha_i' - \alpha_i = 2\tau -2n\pi$
for some $t = \tau$, for each $i$. If either of the $A_i$'s is zero then there
is
only one orbit.
So distinct orbits can be represented by
\beqn
\begin{array}{rclrcl}
q^{\sss 1} &=& A_1\cos(2t-\varphi),&
p_{\sss 1} &=& -A_1\sin(2t-\varphi),\\
q^{\sss 2} &=& A_2\cos(2t),&
p_{\sss 2} &=& -A_2\sin(2t), ~\label{orbits}
\end{array}
\eeqn
where $0 \leq \varphi \leq 2\pi$ and $A_i\neq 0$. (Note that $A_2 = 0$
corresponds to
just {\em one} orbit for all values of $\varphi$.)

Now if an orbit $\cal O$ lies on $\hat\Gamma$ we also have \(A_1^2 + A_2^2 =
R^2\)
so that we can write
\beqn
 A_1 = R\sin\left(\frac{\theta}{2}\right), \quad  A_2 =
R\cos\left(\frac{\theta}{2}\right)
\eeqn
with $0\leq \theta \leq \pi$.
The orbits lying on $\hat\Gamma$ are thus parametrized by the two angles
$\theta \in [0,\pi]$ and $\varphi \in [0,2\pi]$, so that the space of
orbits is $\tilde{\Gamma}=S^2$.

The reduced phase space $\Gamma^*$ is obtained by a gauge choice
$\chi = 0 $ which cuts each of the above orbits once. The surface
$p_2 = 0 $ intersects the orbits (\ref{orbits}) at the points
$t = n\pi /2$ if $A_2 \neq 0$ and at $q_2= 0$ for all $t$ when $A_2=0$.
Now $q_2 =0$ represents one orbit, as discussed earlier. Note that this
is the south pole ($\theta = \pi$) of the space of orbits $\tilde\Gamma$ .
But the other orbits are intersected {\em twice}, i.e. at $q_2 = \pm A_2$,
corresponding to the upper and lower hemispheres of the constraint
surface ${\Gamma ^*}'$.
This means that while the {\em equator} ($q_2 = 0$) maps to the south
pole, {\em both} the hemispheres ($q_2 = \pm A_2$) map to the rest of the
sphere $\tilde\Gamma$: this means we are double-counting. So to get the correct
reduced phase space, we must restrict $q_2$ to be positive (say), so that
$\theta'$
lies in $[0,\pi/2]$ which is the upper hemisphere alone.
 Now we get the reduced phase space as $\Gamma^*=S^2$, on which the Dirac
bracket (\ref{DB})
 actually defines the Poisson bracket--
\beqn
 \{f,g\}^* = {4\over R^2 \sin{\theta}} \left({\partial{f}\over
\partial{\theta}} {\partial{g}\over\partial{\varphi}} -
{\partial{f}\over\partial{\varphi}} {\partial{g}\over\partial{\theta}} \right)
{}~\label{PB}
\eeqn
which is the standard one on a sphere of radius $R/2$, with the
usual coordinate singularity at the poles.
The symplectic structure induced from that on $R^4$ gives the same result
of course.

So the reduced phase space $\Gamma^*$ of the system is $S^2$
after choosing  as the gauge-fixing condition (\ref{c-2}) {\em together}
with the requirement that each gauge orbit is counted as cut only once.

We now proceed to quantize this system by the two methods A and B
outlined in the Introduction.

\section{Method A: Quantization of the Reduced Phase Space}
\setcounter{equation}{0}

We have here a phase space, $S^2$, that is not a cotangent bundle and so
canonical methods of quantization cannot be applied. To quantize this, we
use the technique of geometric quantization \cite{Woodhouse}. A quick
review of this as well as the calculations for $S^2$ are provided in
the appendix.

The sphere is quantizable only if the radius satisfies the Weil integrality
condition:  \quad$R^2 = 2N\hbar, \quad N\in \bf{Z}$ \quad  (cf. equation
\ref{int-cond}
in the appendix).

Working in the complex coordinates $(z,\bar{z})$ obtained by stereographic
projection
through the north pole, the operator corresponding to an observable $f$
satisfying
the quantizability condition (\ref{quant-obs}) is given by
\beqn
\hat{f} &=& {2\hbar\over
R^2}\left(1+|z|^2\right)^2(\partial_{\bar{z}}{f}\partial_z -
\partial_z{f}\partial_{\bar{z}}) \nonumber \\
&-& \left(1+|z|^2\right)\bar{z}\partial_{\bar{z}}{f} + f
\eeqn
(cf. equation \ref{quant-op}).
This acts on an $(N+1)$-dimensional Hilbert space of sections that
are locally given by polynomials in $z$ of order at most $N$.

  There is a natural physical interpretation of this system. The phase
space $S^2$ can be interpreted as describing the classical dynamics
of the spin degrees of freedom of a particle, represented by a vector
${\bf J}$ in ${\bf R}^3$ such that ${\bf J}^2 = j^2$. The magnitude of
${\bf J}$ is preserved and the equations of motion are understood as being
of first order in the time-derivatives.
The components $J_1$, $J_2$ and $J_3$ are given in terms of the
holomorphic coordinates on the sphere of radius $j$ by
\beqn
   	J_1 &=& j{{z+ \bar{z}}\over{1+|z|^2}},\nonumber\\
        J_2 &=&-ij{{z-\bar{z}}\over{1+|z|^2}},\\
        J_3 &=& j{{|z|^2 - 1}\over{1+|z|^2}},\nonumber
\eeqn
and they satisfy the Lie algebra, $\{J_{\sss a},J_b\} = \epsilon_{abc} J_c$, of
$SU(2)$.
 Upon quantizing, the integrality condition (\ref{int-cond}) gives
\beqn
j={N\hbar \over 2}, \quad N\in {\bf{Z}}^+,
\eeqn
and the spin operators are
 \beqn
	\hat{J_1} &=& {\hbar\over 2}[(1-z^{\sss 2})\partial_z + Nz] ,\nonumber\\
\nopagebreak[3]
	\hat{J_2} &=& i{\hbar\over 2}[(1+z^{\sss 2})\partial_z - Nz] ,\\
	\hat{J_3} &=& {\hbar \over 2}[2z\partial_z - N] .\nonumber
   \eeqn

The Hilbert space is ($N+1$)-dimensional.
One can see that $\sum \hat{J_i}^2 = {\hbar^2 \over 4}N(N+2)
= j/\hbar(j/\hbar+1)$.
This is therefore just the standard quantum theory of an
elementary particle with spin $j/\hbar $, which can take
half-integral values. One can
also recover the Pauli matrices as the representation of the
$J_i$'s in the basis $(1,z,\ldots,z^N)$.

\section{Method B: Dirac Quantization }
\setcounter{equation}{0}

The constrained phase space $\Gamma$ is now quantized by the
Dirac method. Of the
two second class constraints (\ref{c-1})
 and (\ref{c-2}), one, in this case $\phi$, is chosen to be the
gauge-generating first class constraint
while the other ($\chi$ in this case) is a gauge-fixing condition
which plays no essential part in this scheme. Now one quantizes ${\bf{R}}^4$
 by the canonical method, i.e., by the association of operators
\beqn
 q^{\sss a}  \rightarrow  \hat{q^{\sss a}} &=& q^{\sss a},\\
 p_{\sss a}  \rightarrow  \hat{p_{\sss a}} &=& {\hbar\over
i}{\partial\over\partial{q^{\sss a}}},
\eeqn
which act on a Hilbert space of square-integrable wave functions
$\Psi(q^{\sss 1},q^{\sss 2})$. Of these only those represent physical states
that are
gauge-invariant. So the operator corresponding to
the gauge constraint must annihilate these state vectors (supplementary
condition):
\beqn
\hat{\phi} \Psi(q^{\sss 1},q^{\sss 2}) = 0  \\ ~\label{g-cond}
\Rightarrow \left[ {q^{\sss 1}}^2 + {q^{\sss 2}}^2 - {\hbar}^2 \left(
{\partial{^2}\over\partial{q^{\sss 1}}^2} + {\partial{^2}\over\partial{q^{\sss
2}}^2}\right)- R^2 \right]
\Psi = 0
\eeqn
which gives
\beqn
\Psi(q^{\sss 1},q^{\sss 2}) = (const)e^{-({q^{\sss 1}}^{\scriptstyle 2} +
{q^{\sss 2}}^{\scriptstyle 2})/2}H_n(q^{\sss 1})H_m(q^{\sss 2})
\eeqn
with
\beqn
R^2 = 2N\hbar,
\eeqn
where $N=n+m+1$ and $n$ and $m$ are non-negative integers.
The radius is thus quantized as even multiples of $\hbar$ and since for each
$R^2 = 2N\hbar$ there are $N$ possible states, the
Hilbert space is $N$-dimensional.

The functions $f(q^{\sss a},p_{\sss a})$ in ${\bf{R}}^4$ that correspond to
 physical observables are those that commute with the gauge
 generator (the so-called first class observables in Dirac's terminology),
i.e.,
\beqn
\{\phi,f\} = 0\nonumber \\
 \Rightarrow\quad q^{\sss 1}{\partial{f}\over\partial{p_{\sss 1}}}
 - p_{\sss 1}{\partial{f}\over\partial{q^{\sss 1}}} + q^{\sss 2}{\partial{f}
\over\partial{p_{\sss 2}}} - p_{\sss 2}{\partial{f}\over\partial{q^{\sss 2}}} =
0.
\eeqn
In the variables $z^{\sss a} = q^{\sss a} + i\delta^{\sss ab} p_{\sss b}$
we have
\beqn
(z^{\sss a}\partial_{z^{\sss a}} - \bar{z}\partial_{\bar{z}^{\sss a}})f(z^{\sss
a},
\bar{z}^{\sss a}) = 0,\\
\Rightarrow \qquad  f(z^{\sss a},\bar{z}^b) = (z^{\sss a})^{k_{\sss
a}}(\bar{z}^b)^{k_b}
\label{D-q-cond}
\eeqn
with \[ \sum_{\sss a} k_{\sss a} = \sum_b k_b .\]
For the corresponding quantum operators to be well-defined,
considerations such as that of self-adjointness may further
restrict this class.

For the sake of comparison with the results of
the previous section, let us look at the $SU(2)$ algebra
generated by the following combinations of quadratic operators
of the type $z^i\bar{z}^j$ :
\beqn
 J_1 &=& \frac{1}{4} (z^{\sss 1}\bar{z}^2 + z^{\sss 2}\bar{z}^1),\nonumber\\
 J_2 &=& \frac{1}{4i} (z^{\sss 2}\bar{z}^1 - z^{\sss 1}\bar{z}^2),\\
 J_3 &=& \frac{1}{4} (z^{\sss 1}\bar{z}^1 + z^{\sss 2}\bar{z}^2).\nonumber
 \eeqn
Quantization, \( J_i \rightarrow \hat{J_i}\) is achieved by \( z^{\sss a}
\rightarrow
\hat{z^{\sss a}}\) and
 \beqn
 \sum_{\sss a} \hat{J_{\sss a}}^2 &=& \frac{1}{16}\left({q^{\sss a}}^2 +
{p_{\sss a}}^2\right)^2 +
\frac{1}{4}(i\hbar)^2 \nonumber\\
 &=& \left(\frac{R^2}{4}\right)^2 - \frac{\hbar^2}{4} \nonumber\\
 &=& {\hbar}^2\left(\frac{N^2}{4} - \frac{1}{4}\right)  \nonumber\\
&=& {\hbar}^2\left(\frac{N}{2} - \frac{1}{2}\right)\left(\frac{N}{2} +
\frac{1}{2}\right).
\eeqn
If this corresponds to ${\hbar}^2j(j+1)$, we get $j = (N-1)/2 $,
the standard result\footnote{ In comparison, geometric quantization
gave an ($N+1$)-dimensional Hilbert space for spin $N/2$. This
slight discrepancy can be rectified by incorporating the
metaplectic correction to geometric quantization(see, for
example,\cite{Tuynman}),
 whereupon the two quantization schemes match exactly.}.

In the present instance, we find that Dirac quantization gives results
equivalent to the quantization of the reduced phase space. In particular,
the quantization of the parameter $R$, which resulted from the Weil
integrality condition in the last section, appears here as a result of the
normalizability of the wave functions.

\subsection{Choice of Constraints}
\label{sec-int_g_cond}
\setcounter{equation}{0}
\renewcommand{\theequation}{\thesection.\thesubsection.\arabic{equation}}

When a system with second class constraints can be regarded as a gauge
theory, half the constraints can be chosen as first class and generate gauge
transformations and the rest are gauge-fixing conditions. When there are only
two second class constraints as in the example considered here, one may choose
either as the gauge generator. The choices may give different theories. Dirac
quantization requires no {\it a priori} criterion for the choice of the
first class (gauge) constraints. Suppose, in the example considered here,
that instead of $\phi$ one chose $\chi$ as the gauge generator and let
$\phi$ be the gauge-fixing condition. The constraint surface ${\Gamma^*}'$
and the singular Dirac brackets remain unaltered. The orbits $\cal {O}$
are in this case the lines $q^{\sss 2}(t) = q^{\sss 2}(0) + t$, and again
the gauge choice $\phi = 0 $ intersects these at two places: $q^{\sss 2}(t)
= \pm q^{\sss 2}(0)$. So the reduced phase space $\Gamma^*$ is again an $S^2$:
the upper hemisphere of ${\Gamma^*}'$ with the (singular) equator mapped
to the south pole; quantization by method A is the same as before.
   However, in Method B, the physical states are
obtained by imposing the constraint condition
\beqn
 \hat{\chi} \Psi(q^{\sss 1},q^{\sss 2}) =0,\nonumber\\
\Rightarrow {\partial\over{\partial{q^{\sss 2}}}}\Psi(q^{\sss 1},q^{\sss 2})
=0,
\eeqn
i.e. $\Psi$ is a function of
$q^{\sss 1}$ alone. The Hilbert space is infinite-dimensional and quantizable
observables are general functions of $q^{\sss 1}, p_{\sss 1}$ and $p_{\sss 2}$.
This quantization is manifestly different from that obtained previously. On
the other hand, suppose we make a different gauge choice: $\chi'\equiv q^{\sss
2} = 0$. $\Gamma^*$ is the surface $q^{\sss 2} = p_{\sss 2} =0$, which is $R^2$
and the Dirac quantization discussed in the beginning of the section gives
the usual quantization on this, so that methods A and B give equivalent
results. So here we see that method A gives different quantizations
for different gauge choices.

This source of this `discrepancy' can be traced to the observation made
in the introduction regarding the intersection of the gauge-fixing surfaces
with the gauge orbits. In the first case considered here,
we carefully obtained the reduced phase space as $S^3/S^1 = S^2$, and chose
a gauge that selected one 2-sphere for every $R$.
 In the second case, though we were careful in considering only one
intersection of the surfaces generated by the gauge choice $ \phi =
0 $ with the gauge orbits $\cal O$, {\em not all orbits are cut.} A gauge
 choice that intersects {\em all} the gauge orbits is $\chi' = 0$.
 So in the former case, one was artificially truncating the true
 phase space by an inappropriate gauge choice, and thereby obtained a
different dynamical system.

  Now the Dirac quantization method makes no reference to any gauge-fixing
 and is determined once the gauge generators ($\phi_i$'s)
 are specified and a supplementary condition is imposed to ensure that the
Hilbert space so constructed is associated with the true phase space
$\tilde{\Gamma}$.
Quantization of the reduced phase space $\Gamma^*$ can be expected to give
equivalent
results only when $\Gamma^*$ is diffeomorphic (symplectically) to
$\tilde{\Gamma}$.
So the gauge-fixing constraints $\chi_i$ must satisfy not only the condition
det$\{\phi_i,\chi_j\} \neq 0$, which ensures the selection of one point
from each gauge orbit {\em locally}, but also that the resultant reduced
phase space $\Gamma^*$ be diffeomorphic to the  space $\tilde{\Gamma}$
of orbits. This point may seem obvious in retrospect but in practice
one may miss it get a resultant quantum theory which may be consistent
but not reflect gauge-independence.

\section{Discussion and Conclusions}

  In the context of a simple gauge theory, viz. $S^2$ as a phase space,
we have analyzed and compared two methods of quantization, viz. quantization of
the reduced phase space and Dirac quantization, and examined a condition
for their equivalence.

 Another observation refers to quantizability itself.
As is well known, in geometric quantization there exists a condition on
the phase space for quantizability: the Weil integrality condition must
be satisfied if the pre-quantum bundle is to exist. In the case of
$S^2$ this restricts the radius to discrete values. This slightly
counter-intuitive result is not merely a peculiarity of the geometric
approach. As shown in the present example, this quantizability
condition reappears though in a different guise-- it is a result of
the physical Hilbert space being well defined (square-integrability of the wave
functions). This shows that the quantizability condition is related to the
global topological properties of the phase space.

  The comparison of quantizable observables shows that there exists
a restricted class of classical observables that can be consistently
quantized. Conditions of self-adjointness and operator commutativity
with the gauge generators must also hold rigorously.

 Gauge theories are encountered in many contexts and in particular cases,
either method of quantization may prove convenient. We have not considered
the possible difficulties in applying either of these methods, but assuming
they have been tided over, one needs to be careful about capturing the
true phase space of the Dirac approach (method B) in the reduced phase
space of method A. Further, restrictions may
be encountered in parameters entering the theory via the constraints,
for example $R$ in the present example.

\section*{Appendix}
\setcounter{equation}{0}
\renewcommand{\theequation}{A.\arabic{equation}}

This is a brief review of geometric quantization and its application
to the quantization of $S^2$.

The classical phase space $\Gamma$ is a $2n$-dimensional symplectic
manifold. The symplectic form $\omega$  defines a Poisson
algebra $\cal{A}$ of observables, which are $C^\infty$ functions on
 $\Gamma$.
 In formulating a quantization, i.e. a map from $\cal{A}$ to the set
 $\cal{Q}$ of operators acting on
a Hilbert space $\cal{H}$, the basic guidelines were spelt out by Dirac:
\begin{itemize}
\item[$\bullet$]  The map $\cal{A} \rightarrow \cal{Q} $ is linear.

\item[$\bullet \bullet$] Constants are mapped to multiples of the identity
operator.

\item[$\bullet \bullet \bullet $] For classical observables $f_i\in \cal{A}$
and the corresponding quantum operators $\hat{f}_i \in \cal{Q}$,
      \begin{center}
      $ [\hat{f_1},\hat{f_2}] = k\hat{f_3} ,\qquad f_3 = \{f_1,f_2\}$
      \end{center}
      where \{.,.\} is the Poisson bracket and [.,.] is the commutator. k is
      some constant, canonically $i\hbar$.
\end{itemize}

Geometric quantization typically achieves this in two stages.
 The first stage, called `prequantization'
 involves finding such a map.
The prequantum Hilbert space is, however, too large to be a
physically reasonable quantum description. The wave functions
depend on all the phase space variables, so that the standard
 Schr\"{o}dinger description is not obtained even in the case
of ${\bf R}^{2N}$. Also, group
representations of elementary systems turn out to be reducible.
Hence we need stage two of geometric quantization, which is the
 choice of a polarization of the manifold.

The `prequantum' operator
corresponding to $f\in C^\infty(\Gamma)$ is constructed as follows:
 in local
Darboux coordinates, $(q^{\sss a}, p_{\sss a})$,
\beqn
 \omega &=& dp_{\sss a}\wedge dq^{\sss a}  \\
 &=& d(p_{\sss a}dq^{\sss a})
\eeqn
 so that the symplectic potential is
\beqn
\theta = p_{\sss a}dq^{\sss a} .
\eeqn
The Hamiltonian vector field $X_f$ corresponding to $f$,
is
\beqn
  X_f&=& {\partial{f}\over {\partial{p_{\sss a}}}}\partial_{q^{\sss
a}}-{\partial{f}\over {\partial{q^{\sss a}}}}\partial_{p_{\sss a}} \eeqn
and
\beqn
\theta(X_f) = p_{\sss a}{\partial{f}\over\partial{p_{\sss a}}}.
\eeqn
Then the operator representation of $f$ is
\begin{eqnarray}
	\hat{f} = -i\hbar X_f  + \theta(X_f)  + f.
\end{eqnarray}
This acts on sections of a complex line bundle
$\cal{B}$ over $\Gamma$, the connection potential on which is
$\theta/\hbar$ (the curvature being $\omega/\hbar$).
 A compatible Hermitian structure $(.,.)$ must be defined on it. Now,
\begin{eqnarray}
	\hat{f} = -i\hbar \nabla_{X_f} + f .
\end{eqnarray}
where $\nabla_{X_f} = X_f - {i\over\hbar} \theta(X_f)$.
Such a
 line bundle exists if and only if $\omega$ satisfies the
Weil integrality condition \cite{Weil}. This might restrict the
class of classical phase spaces that can be quantized in this
approach. One way of stating this condition, for a simply connected
 manifold, is that the integral
 of $\omega/\hbar$ over any closed, oriented  2-dimensional
submanifold of $\Gamma$ is an integral multiple of $2\pi$.

Stage two of geometric quantization involves the choice of a polarization
$P$ of the manifold. Sections of $\cal{B}$
constant along the polarization form the quantum Hilbert space $\cal{H}_{Q}$.
If the vector fields $X_m$ are tangent to $P$ at a point
$m$ on $\Gamma$, then
a section $s:\Gamma\rightarrow\cal{B} $ is said to be polarized if
\begin{eqnarray}
\nabla_{X_m} s = 0 .
\end{eqnarray}
Thus $s$ is a function of only $n$ variables.
So if $(s,s)$ is the Hermitian product on $\cal{B}$, the Hilbert space
 $\cal{H}_{Q}$ consists of polarized sections $s$ such that
\begin{eqnarray}
	\langle s,s\rangle = \int_{\Gamma} (s,s)\,\omega^n\; < \infty.
\end{eqnarray}

In this scheme, only those observables can be directly quantized that
 preserve the polarization : if $s$ is a polarized section, so must
 be $\hat{f}s$, meaning if $X$ is a vector field tangent to $P$
then we must have
\[ \nabla_X(\hat{f}s) = f(\nabla\!_X \,s) - i\hbar\nabla_{[X,X_f]} s = 0 ,\]
i.e. $ [X,X_f]$ must also be tangent to $P$.

Further refinements such as half-density quantization and metaplectic
corrections are not considered here as this level is sufficient for
the case in hand.

This is applied to $S^2$ in the following.

The 2-sphere is a symplectic manifold on which the measure,
in spherical coordinates $(\theta,\phi)$,
\beqn
	\omega= \frac{R^2}{4}\sin\theta d\theta\wedge d\phi \label{w}
\eeqn
(where $R/2 $ is the radius) serves as the symplectic form corresponding
to the Poisson bracket (\ref{PB}). It is more convenient to look upon $S^2$ as
  a K\"{a}hler manifold with holomorphic coordinates
$z_i$ obtained by stereographic projection through the north (south)
 poles:
\[ z_n = \cot\left(\frac{\theta}{2}\right) e^{i\phi}, \qquad
z_s =\tan\left(\frac{\theta}{2}\right) e^{-i\phi} .\]
So it is covered by two charts $U_n$ and $U_s$ both isomorphic to the complex
plane {\bf{C}}.
Note that  $z_n = 1/z_s$.
 Working in the northern chart, $z\in U_n \approx \bf{C} $,
the symplectic form is
\begin{eqnarray}
 \omega_n = -i\frac{R^2}{2} \left(1+|z|^2\right)^{-2}\:dz\wedge d\bar{z}.
{}~\label{sympl}
\end{eqnarray}
The symplectic potential is
\beqn
\theta_n = -i\frac{R^2}{2}\left(1+|z|^2\right)^{-1}\: \bar{z}dz .
\eeqn

The prequantum line bundle ${\cal B}$, which is locally $U_i\times\bf{C}$, must
 have a curvature $\omega/\hbar$. This means that the transition
function $c_{ns}$ (on the overlap $U_n\cap U_s$) must be given by
\beqn
 \theta_s - \theta_n = i\hbar\: d\ln c_{ns} ,
\eeqn
which gives
\beqn
  c_{ns} = z^{R^2/2\hbar} .
\eeqn
This is well defined only for $R^2/2\hbar \in \bf{Z}$.
This of course is the Weil integrality condition :
\beqn
 \int_{S^2 }\omega/\hbar = 2N\pi \quad \Rightarrow \quad
R^2 = 2N\hbar, \quad  N\in \bf{Z}. ~\label{int-cond}
\eeqn
The Hamiltonian vector field corresponding to a function $f(z,\bar{z})$ is
\begin{eqnarray}
 X_f = {2i\over R^2} \left(1 + |z|^2\right)^2(\partial_{\bar{z}}
{f}\partial_z - \partial_z{f}\partial_{\bar{z}}),
\end{eqnarray}
and\beqn
\theta  (X_f) = \left(1+|z|^2\right)\bar{z}
\partial_{\bar{z}}{f}.\label{eq-pot}
\eeqn
So the quantum operator corresponding to $f$ is
\beqn
\hat{f} &=& {2\hbar\over
R^2}\left(1+|z|^2\right)^2(\partial_{\bar{z}}{f}\partial_z
 - \partial_z{f}\partial_{\bar{z}}) \nonumber \\
&-& \left(1+|z|^2\right)\bar{z}\partial_{\bar{z}}{f} + f.
\label{quant-op}
\eeqn

A natural choice of polarization  is the K\"{a}hler polarization \footnote{On
$S^2$ there exist no real polarizations.}
spanned by the Hamiltonian vector fields generated by the
holomorphic coordinates
\beqn
X_z = -{2i\over R^2}\left(1+|z|^2\right)^2\partial_{\bar{z}}.~\label{pol}
\eeqn
There then exists a scalar $K$ in the neighbourhood of each point such
that the symplectic potential  $\theta$ given by (\ref{eq-pot}) can
 be expressed as
\beqn
\theta = -i\partial_z{K}, ~\label{k}
\eeqn
where $ K = \frac{R^2}{2}\ln\left(1 + |z|^2\right)$.
This potential annihilates the vectors (\ref{pol}) i.e., $\theta(X_z) = 0$,
 and is said to be adapted to the polarization.
 The polarized sections of ${\cal B}$ satisfy
\beqn
\nabla_{X_z}\, s(z,\bar{z}) &  =  & 0 ,\\
\Rightarrow \; \partial_{\bar{z}}s & = & 0 .
\eeqn
So $\cal{H}_Q$ consists of holomorphic sections of ${\cal B}$ . In order
that the wave functions be well defined on all of $S^2$, they must be well
 defined in the overlap region $U_n \cap U_s$, where they are related by
\[ \psi_n = c_{ns}\psi_s \]
i.e., \beqn
 \psi_s\left(\frac{1}{z}\right) = z^{-N}\psi_n(z).
\eeqn
So $\psi(z)$ must be a polynomial in $z$ of order at most $N$.
$\cal{H}_Q$ is therefore spanned by the set $(1,z,\ldots,z^N)$
and is $(N+1)$-dimensional. Given the scalar $K$ of (\ref{k}),
the Hermitian structure on $\cal{B}$ can then be chosen to be
\beqn
 (s,s)= \bar{s}s e^{-K/\hbar}.
\eeqn
So the inner product on the Hilbert space is given by
\beqn
 \langle\psi,\psi\rangle\, &=& \int_{\bf{C}} \bar{\psi}\psi
\left(1+|z|^2\right)^{-N}\, \omega \nonumber\\
 &=& \int_{\bf{C}} \bar{\psi} \psi \left(1+|z|^2\right)^{-N-2} dz d\bar{z}.
\eeqn
A similar result holds for the chart $U_s$.

Quantizable observables $f(z,\bar{z})$ in this scheme must satisfy
$[X_z,X_f] \in P$.
Now
\beqn
	[X_z,X_f] = {2i\over R^2}{\partial \over \partial{z}}
\left(\;\left(1+|z|^2\right)^2{\partial{f}\over\partial{\bar{z}}}\;\right)X_z
\nonumber \\
+ {2i\over R^2}
{\;\partial\over\partial{\bar{z}}}\left(\;\left(1+|z|^2\right)^2{\partial{f}\over
\partial{\bar{z}}}\;\right)X_{\bar{z}}.
\eeqn
For this to belong to the polarization, the second term must vanish, i.e.,
\beqn
{\partial\over\partial{\bar{z}}}\left(\;\left(1+|z|^2\right)^2\;
{\partial{f(z,\bar{z})}\over\partial{\bar{z}}}\right) = 0.\\
\Rightarrow \; f(z,\bar{z}) = {{h_1(z)+ \bar{z} h_2(z)}\over {1+|z|^2}}.
{}~\label{q-cond}
\eeqn
where $h_1$ and $h_2$ are functions of $z$ alone.
 For such an observable to be well-defined on all of $S^2$,
one requires that it have the same value in both charts:
\[ {{h_1(z)+ \bar{z} h_2(z)}\over {1+|z|^2}} =  {{h'_1(z')+ \bar{z}'
h'_2(z')}\over {1+|z'|^2}} \]
which gives

\[  h'_1(z) = z h_2(1/z) \]
\[  h'_2(z) = z h_1(1/z)\]
where the prime denotes the southern chart. These must be well defined
for all $z$.
Further, restriction to real functions alone gives a general observable
the form:
\beqn
  f(z,\bar{z}) = {{a + bz + \bar{b}\bar{z} + c|z|^2}\over{1 + |z|^2}}
{}~\label{quant-obs} \eeqn where the constants $a$ and $c$ are real and $b$
is complex. Dynamics in this theory can be dictated by a Hamiltonian
chosen from this
 class of observables. This completes the quantization of $S^2$.

\section*{Acknowledgements}

I am indebted to G.Date for suggesting this problem and thank
him for fruitful discussions. I would also like to thank A.P.Balachandran for
discussions.


\begin{thebibliography}{20}

\bibitem{Dirac} P.A.M.Dirac, {\em Lectures on Quantum Mechanics},
Yeshiva University (1964);\\
M.Henneaux and C.Teitelboim, {\em Quantization of Gauge Systems},
Princeton University Press (1992).

\bibitem{Woodhouse} N.M.J.Woodhouse, {\em Geometric Quantization.}
2nd edition, Clarendon, Oxford (1992).

\bibitem{Weil} see ref \cite{Woodhouse} above or Simms and Woodhouse, {\em
Lectures on
 Geometric Quantization}, Lecture Notes in Physics 53 (1976).

\bibitem{Tuynman} G.M.Tuynman,``Generalized Bergman Kernels and
Geometric Quantization'', J.Math.Phys. {\bf 28}(3) 573 (1987).

\end{thebibliography}
\end{document}